\documentclass[twocolumn,showpacs,preprintnumbers,amsmath,amssymb]{revtex4}

\usepackage{graphicx}
\usepackage{dcolumn}
\usepackage{bm}
\usepackage{times,mathptm,amsmath,amssymb,graphicx,here,fancyhdr} 

\begin{document}


\title{Efficiency in nanostructured thermionic and thermoelectric devices}

\author{M.~F. O'Dwyer}
\email{mo15@uow.edu.au}
\author{R.~A. Lewis}
\author{C. Zhang}
 \affiliation{School of Engineering Physics and Institute for Superconducting and Electronic Materials,
 University of Wollongong, Wollongong NSW 2522, Australia}

\author{T.~E. Humphrey}
\affiliation{School of Physics, University of New South Wales,
Sydney NSW 2052, Australia\\
Baskin School of Engineering, University of California Santa Cruz,
Santa Cruz, CA 95064-1077}

\date{\today}

\begin{abstract}
Advances in solid-state device design now allow the spectrum of
transmitted electrons in thermionic and thermoelectric devices to be
engineered in ways that were not previously possible. Here we show
that the shape of the electron energy spectrum in these devices has
a significant impact on their performance. We distinguish between
traditional thermionic devices where electron momentum is filtered
in the direction of transport only and a second type, in which the
electron filtering occurs according to total electron momentum. Such
`total momentum filtered' $k_{r}$ thermionic devices could
potentially be implemented in, for example, quantum dot
superlattices. It is shown that whilst total momentum filtered
thermionic devices may achieve efficiency equal to the Carnot value,
traditional thermionic devices are limited to efficiency below this.
Our second main result is that the electronic efficiency of a device
is not only improved by reducing the width of the transmission
filter as has previously been shown, but also strongly depends on
whether the transmission probability rises sharply from zero to full
transmission. The benefit of increasing efficiency through a sharply
rising transmission probability is that it can be achieved without
sacrificing device power, in contrast to the use of a narrow
transmission filter which can greatly reduce power. We show that
devices which have a sharply-rising transmission probability
significantly outperform those which do not and it is shown such
transmission probabilities may be achieved with practical single and
multibarrier devices. Finally, we comment on the implications of the
effect the shape of the electron energy spectrum on the efficiency
of thermoelectric devices.
\end{abstract}



\maketitle

\section{Introduction}

Traditional vacuum thermionic power generators
\cite{SchilichterPhD1915, WilsonJAP1959, HoustonJAP1959} with
macroscopic gaps between emitter and collector plates are limited to
very high temperature applications ($T_{H} >$ 1000~K). Refrigeration
using such devices, as first suggested by Mahan \cite{MahanJAP1994},
is also limited to high temperatures due to a lack of suitable
materials with work-functions below $\sim$0.3~eV.

Nanostructures are currently being investigated in an attempt to
develop thermionic devices that can refrigerate or generate power at
lower temperatures. The potential for achieving lower barrier
heights via the use of semiconductor heterostructures was pointed
out by Shakouri and Bowers \cite{ShakouriICT1997, ShakouriAPL1997},
with Mahan et al. \cite{MahanPRL1998, MahanJAP1998} suggesting
multilayers as a means of reducing the phonon heat leaks inherent in
the use of solid-state rather than vacuum devices. Successful
solid-state thermionic cooling of up to a few degrees has been
reported
\cite{ShakouriAPL1999,LabountySPIE2000,FanEL2001,FanAPL2001}.

Another promising direction is the use of nanometer gaps between the
emitter and the collector to lower the work function via quantum
tunnelling \cite{HishinumaAPL2001, HishinumaAPL2002}, with Hishinuma
et al. reporting cooling of about a mK in such a system. Cooling by
field emission from carbon nanotubes and other nanostructures has
also been proposed \cite{MiskovskyAPL1999, FisherJHT2002}.

In thermoelectrics, nanostructured devices may offer the possibility
of substantially increasing the thermoelectric figure of merit,
$ZT$, over that of traditional bulk bismuth telluride based devices
($ZT \approx 1$) due to enhanced electron transport and phonon
blocking properties. Hicks and Dresselhaus have predicted that $ZT$
can be enhanced using quantum-well superlattices
\cite{HicksPRB1993a} and quantum wires \cite{HicksPRB1993b}.
Venkatasubramanian et al. have reported the highest $ZT$ to date,
$ZT\sim2.4$ using a p-type Bi$_{2}$Te$_{3}$/Sb$_{2}$Te$_{3}$
superlattice \cite{VenkatasubramanianNature2001}. Other methods used
and suggested for the enhancement of the figure of merit include the
use of quantum-dot superlattices
\cite{HarmanScience2002,BalandinAPL2003}, superlatices with a
non-conservation of lateral momentum \cite{VashaeeJAP2004,
VashaeePRL2004}, inhomogeneous doping \cite{HumphreyPRL2005} and
nanotubes \cite{ZhaoAPL2005}.

Many of these approaches offer the possibility of engineering the
energy spectrum (the number of electrons transmitted through the
device as a function of energy) in a way that was not possible in
traditional vacuum thermionics or bulk thermoelectrics. In light of
the new design freedom offered by nanostructures, it is useful to
re-examine the impact of the electron energy spectrum upon what has
been called the `electronic efficiency' of thermionic devices
\cite{Hatsopoulos1973}, defined as the efficiency associated with
strictly electronic processes under ideal conditions of particle
transport.

Improvements in electronic efficiency due to better device design
which can be achieved without lowering the power will translate into
an improvement in the operating efficiency of practical thermionic
devices where non-ideal effects, such as phonon and radiative leaks,
as well as contact and lead resistances, are important. To achieve
high overall efficiency in practical devices it is important to
design devices that not only achieve low thermal conductivity, but
high electronic efficiency at finite power as well.

\begin{figure}
\includegraphics[width=2.7in]{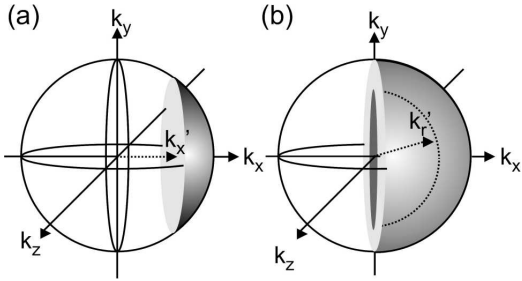}
\caption{\label{momentum_space} Fermi spheres indicating electrons
transmitted by (a) a $k_{x}$ filtered device and (b) a $k_{r}$
filtered device.}
\end{figure}

In this paper we analyse in detail the dependence of the electronic
efficiency of thermionic power generators and refrigerators upon the
details of the energy spectrum of electrons transmitted
ballistically between the emitter and collector. The term energy
filtering is often used to indicate a restriction of electron flux
to all those electrons above a certain energy. Here the term energy
filtering will be used in a more general sense to indicate any
arbitrary restriction on the energy spectrum of transmitted
electrons. We examine two idealised models of thermionic
nanodevices. In the first, energy (or more precisely, momentum)
filtering of electrons occurs in the direction of transport only.
This model, which we will denote as a `$k_{x}$-filtered thermionic
device', is applicable to single-barrier or multibarrier
(superlattice) solid-state devices. In the second model, which we
will denote as a `$k_{r}$-filtered thermionic device', energy
filtering of the total energy of electrons is assumed to be
possible. This model is applicable to vacuum emission from
nanostructures such as carbon nanotubes and solid-state devices in
which there is periodic modulation of the potential in all three
dimensions (such as quantum dot superlattices), or superlattices in
which there is non-conservation of electron momentum in directions
perpendicular to transport \cite{VashaeeJAP2004, VashaeePRL2004}.
Fig. \ref{momentum_space} shows geometrically the range of electrons
transmitted in idealised $k_{x}$ and $k_{r}$ type devices in
momentum space.

With thermoelectric devices, the presence of a bandgap in different
crystallographic directions ensures electrons contributing to the
current have a certain minimum value of momentum in all three
dimensions. The close relationship between thermionic and
thermoelectric devices has been analysed by a number of authors
\cite{MahanJAP1998,NolasJAP1999,ViningJAP1999,HumphreyJAP2005}. In
this paper, this comparison is extended to similarities between the
dependence of the electronic efficiencies of these devices on the
details of the electron energy spectrum.

Three main results are presented in this paper. Firstly, it is shown
that while $k_{r}$ devices may achieve electronic efficiency equal
to the Carnot value, conventional $k_{x}$ devices are fundamentally
limited to efficiencies less than this. Secondly, the details of the
electron energy spectrum are shown to have a significant impact on
the electronic efficiency of the device. Narrower electron energy
filters and, more significantly, a sharp rise in the transmission
probability from zero to complete transmission give dramatic
improvements in electronic efficiency. Using a numerical model of
ballistic transport in semiconductor hetrostructures, we show that
sharply-rising transmission probabilities yielding high electronic
efficiency and power may be achieved with wide single barrier and
multibarrier devices. Finally, the equivalence of the ballistic and
diffusive formalisms for devices with length the order of the
electron mean free path \cite{HumphreyJAP2005} means that in this
regime, the electronic efficiency of thermionic and thermoelectric
devices will have the same dependence upon the details of the
electron energy spectrum.

\section{Transport Theory}

\subsection{Ballistic Transport Theory}

A thermionic device consists of two electron reservoirs at different
temperatures and electrochemical potentials, separated by a barrier,
or series of barriers, which limit the flow of electrons between
them to a certain energy range. Whether the device operates as a
power generator, pumping high-energy electrons from the hot to the
cold against the electrical potential difference, or as a
refrigerator, removing high-energy electrons from the cold
reservoir, depends upon the relative magnitudes of the opposing
temperature and electrochemical potentials.

In a $k_{r}$-filtered thermionic device, where the transmission
probability, $\zeta$, is a function of the total electron energy, $E
= \hbar^{2} \textbf{k}^{2} / 2m^{*}$, the net electrical current
density flowing from the cold to the hot reservoir is
\begin{equation}
    \label{kr_electrical}
    J_{r} = q \int^{\infty}_{0} \left[ n^{C}_{r} - n^{H}_{r}
    \right] \zeta(E) dE
\end{equation}
where $q = -1.602 \times 10^{-19}$ C is the charge of an electron
and
\begin{equation}
    \label{kr_flux}
    n_{r}^{C/H} = \dfrac {m^{*}E}{2\pi^{2}\hbar^{3}} f(E,\mu_{C/H},T_{C/H})
\end{equation}
is the number of electrons with total energy $E$ arriving at the
three-dimensional reservoir interface per unit area per unit time
and
\begin{equation}
    \label{Fermi-Dirac}
    f(E,\mu_{C/H},T_{C/H}) = \left[ 1 + \exp \left( \dfrac{E - \mu_{C/H}}{k_{B}T_{C/H}} \right) \right] ^{-1}
\end{equation}
is the Fermi-Dirac distribution function in the cold/hot reservoir
with chemical potential $\mu_{C/H}$ and temperature $T_{C/H}$. We
have assumed that electron velocity is determined by the reservoirs.
A more detailed theory would be required to account for any velocity
changes due to the device structure.

One may calculate the heat flux out of the hot and cold reservoirs
by noting that an electron leaving or entering the cold/hot
reservoir will remove or add respectively an amount of heat equal to
the difference between the energy of the electron and the average
energy of electrons in the reservoir, that is $E - \mu_{C/H}$.
Introducing this factor inside the integral for number current we
may obtain expressions for the net heat flux out of the cold/hot
reservoir as
\begin{equation}
    \label{kr_heat}
    \dot{Q}^{C/H}_{r} = \mp \int^{\infty}_{0} (E - \mu_{C/H} )\left[ n^{C}_{r} - n^{H}_{r}
    \right] \zeta(E) dE.
\end{equation}

In a $k_{x}$-filtered device the transmission probability is a
function of what may be loosely defined as the `kinetic energy of
electrons in the $x$ direction', $E_{x} = \hbar^{2} k_{x}^{2} /
2m^{*}$. It is therefore convenient in this case to write the
electrical and heat currents in terms of $E_{x}$ \cite{Davies1998}
\begin{equation}
    \label{kx_electrical}
    J_{x} = q \int^{\infty}_{0} \left[ n^{C}_{x} - n^{H}_{x}
    \right] \zeta(E_{x}) dE
\end{equation}
where
\begin{equation}
    \label{kx_flux}
    n_{x}^{C/H} = \dfrac {m^{*}k_{B}T_{C/H}}{2\pi^{2}\hbar^{3}} \log
    \left[ 1 + \exp \left( - \dfrac{E_{x} - \mu_{C/H}}{k_{B}T_{C/H}} \right)
    \right]
\end{equation}
is the number of electrons with kinetic energy in the $x$ direction
$E_{x}$ arriving at the reservoir interface per unit area per unit
time.

In a $k_{x}$-filtered device, the average heat removed from the
cold/hot reservoir when an electron with energy in the $x$ direction
$E_{x}$ leaves, $E_{x} + k_{B}T_{C/H} - \mu_{C/H}$ (assuming
Maxwell-Boltzmann statistics), is not the same as that added when an
electron with energy in the $x$ direction $E_{x}$ arrives, $E_{x} +
k_{B}T_{H/C} - \mu_{C/H}$. This difference is due to the fact that,
while the barrier system filters electrons according to their
momentum in the direction of transport, their momenta in the other
two dimensions may take any value, contributing on average an extra
$k_{B}T_{C/H}$ to the energy of electrons emitted from the cold/hot
reservoir. The heat flux out of the cold/hot reservoir in a $k_{x}$
filtered device is therefore given by
\begin{eqnarray}
    \label{kx_heat}
    \dot{Q}^{C/H}_{x} =  \mp \int^{\infty}_{0} \left[ (E_{x} +
    k_{B}T_{C/H} - \mu_{C/H} ) n^{C}_{x} \right. \nonumber \\
    \left. - (E_{x} + k_{B}T_{H/C} - \mu_{C/H} ) n^{H}_{x}
    \right] \zeta(E_{x}) dE.
\end{eqnarray}

It may be noted that many cryogenic ballistic refrigerators such as
normal-insulating-semiconductor (NIS) junction devices
\cite{NahumAPL1994, LeivoAPL1996, ManninenAPL1997} and quantum dot
refrigerators \cite{EdwardsPRB1995} utilise either two- or
one-dimensional reservoirs where the difference between $k_{x}$ and
$k_{r}$ filtered devices are less dramatic or non-existent.

The electronic efficiency as a power generator and coefficient of
performance (COP) as a refrigerator for both $k_{x}$ and $k_{r}$
filtered devices are given by
\begin{equation}
    \label{efficiency}
    \eta^{PG} = VJ / \dot{Q}^{H}
\end{equation}
and
\begin{equation}
    \label{COP}
    \eta^{R} = \dot{Q}^{C} / VJ
\end{equation}
respectively, where $V = (\mu_{C} - \mu_{H})/q$.

\subsection{Diffusive Transport Theory}
\label{Diffusive_Transport_Theory}

Thermoelectric devices are generally differentiated from thermionic
devices according to whether electron transport is diffusive or
ballistic \cite{MahanJAP1998}. There is, however, little to
distinguish the underlying thermodynamics of the two types of
device, with both achieving reversibility under the same conditions
\cite{HumphreyPRL2002, HumphreyPRL2005} and both being governed by
the same `materials parameter'
\cite{ViningJAP1999,UlrichJAP2001,HumphreyJAP2005}.

Under the relaxation-time approximation the electric current in a
thermoelectric device may be calculated using the Boltzmann
transport equation as
\begin{equation}
    \label{diffusive}
    J^{d} = \iiint q D_{l} \left[ v^{l}_{x} \right]^{2} \tau \dfrac{df}{dx} d\textbf{k}.
\end{equation}
where $D_{l}$ is the local density of states (DOS), $\tau =
\tau_{0}E^{r}$ is the relaxation time, and $v^{l}_{x} =
(1/\hbar)[\partial E(k_{x}) /
\partial k_{x}]$ is the velocity in the direction of transport. The
electron energy spectrum in a diffusive device is thus determined by
$D_{l} \left[ v^{l}_{x} \right]^{2} \tau [df/dx]$.

The transport equation for ballistic devices, where the mean free
path of an electron between collisions is greater than the width of
the barrier, or system of barriers, may be written similarly as
\begin{equation}
    \label{ballistic}
    J^{b} = \iiint q D_{r} \zeta v^{r}_{x} \Delta f d\textbf{k}
\end{equation}
where $D_{r} = 1 / (2\pi)^{d}$ is the DOS in $k$-space in the
reservoirs where $d$ is the dimensionality of the reservoirs, and
$\Delta f = f_{C} - f_{H}$ is the difference between the
distribution functions in the cold and hot reservoirs. The electron
energy spectrum in a thermionic device is therefore determined by
$D_{r} v^{r}_{x} \zeta \Delta f$.

We expect that Eqs. \ref{ballistic} and \ref{diffusive} should yield
the same results for devices of width close to the electron mean
free path. If we take the energy dependence of the relaxation time
to be $r = -1/2$, which is appropriate when scattering is dominated
by acoustic phonons, the mean free path in the direction of
transport will be independent of energy and given by $\lambda \equiv
v_{x} \tau$ \cite{HumphreyJAP2005}. For a small piece of
thermoelectric material of length approximately equal to the
electron mean free path $df/dx \approx \Delta f / \lambda$. Eq.
\ref{diffusive} then reduces to \cite{HumphreyJAP2005}
\begin{equation}
    \label{diffusive_MFP}
    J^{d} = \iiint q D_{l} v^{l}_{x} \Delta f d\textbf{k}
\end{equation}
and is of the same form as that of the ballistic transport equation,
Eq. \ref{ballistic}. Thus, the term $D_{l} v^{l}_{x}$ in the
diffusive formalism plays the same role as $D_{r} v^{r}_{x} \zeta$
in the ballistic formalism. We therefore expect the dependencies of
the electron energy spectrum in both thermionic and thermoelectric
device to be similar. Since $v^{r}_{x}$ and $D_{r}$ are fixed by the
reservoirs, at fixed temperature/chemical potential the electron
energy spectrum in a ballistic device is determined by the
transmission probability as device structure varies. In a diffusive
device, both $D_{l}$ and $v^l_x$ may change when the device
structure is altered and affect the energy spectrum.

The heat-current density out of the cold/hot reservoir is given by
\begin{equation}
    \label{TE_heat}
    \dot{Q}_{C/H} = \mp \iiint (E - \mu_{C/H}) D_{l} v^l_{x} \tau \dfrac{df}{dx}
    d\textbf{k}.
\end{equation}

Due to the equivalence of the diffusive and ballistic formalisms in
this regime, the intensive efficiency across a small section of
thermoelectric material \cite{SnyderAPL2003,ViningSTM1997} and the
electronic efficiency/COP of a ballistic device are given by Eqs.
\ref{efficiency} and \ref{COP} respectively.

\section{Reversible Electron Transport}

To achieve reversibility in a thermionic or thermoelectric device,
electrons must flow only at energies where the Fermi occupation of
states, Eq. \ref{Fermi-Dirac}, is constant
\cite{HumphreyPRL2002,HumphreyPRL2005}. Assuming a finite
temperature difference at each end of the device, there are two
different quasi-static limits in which this requirement is
satisfied.

The first way is to restrict the flow of electrons to those with
energies approaching infinity where the occupation of states tends
to zero. This may be achieved, for instance, with an intrinsic
semiconductor where the band gap approaches infinity in a
thermoelectric device or an infinitely high barrier system in a
thermionic device. In vacuum thermionic devices operating at very
high temperatures ($T_{H} > 1500$ K), optimim efficiency is
approached when the barrier height is almost 20 times larger than
$k_{B}T_{H}$ \cite{Hatsopoulos1973}, meaning that electronic
efficiencies close to the Carnot limit may be obtained. However, at
the more moderate emitter temperatures, 300 K $< T_{H} < 800$ K, of
interest in most applications, achieving finite power production or
refrigeration via a thermionic device requires a much lower barrier
height, of the order of a few $k_{B}T_{H}$. It is therefore more
practical to utilize the other quasi-static limit to achieve high
electronic efficiencies.

The second way to achieve reversibility in a thermionic or
thermoelectric device, which we refer to as energy-specific
equilibrium, is to allow electrons to flow only at a single energy
where the Fermi occupation of states throughout the device is the
same \cite{HumphreyPRL2002},
\begin{equation}
    \label{E0}
    E_{0} = \dfrac { \mu (T + \delta T) - (\mu - \delta \mu) T }{ \delta T
    }
\end{equation}
where $\delta T$ and $\delta \mu$ are the temperature and chemical
potential changes respectively over a distance $\delta x$ in the
device. At this energy, the effect, or, in the language of
irreversible thermodynamics, the `affinity' \cite{Callen1960}, of
the opposing temperature and electrochemical potential gradients
upon electrons exactly cancels and transport occurs reversibly. This
is also the energy at which the energy-resolved current changes
sign, that is, for electrons with energies less than $E_{0}$ the net
current flows from the hot to cold reservoir and for energies
greater than $E_{0}$ net current flows from the cold to hot
reservoir. Transport of electrons of a single energy only might be
achieved using resonant tunneling in a superlattice or quantum dot
superlattice. For a thermionic device, the ballistic transmission
energy is determined substituting the cold and hot reservoir
temperatures and chemical potentials into Eq. \ref{E0}. Here we
shall denote a filtering system which transmits only a single energy
of electrons between the reservoirs, be that the single total energy
for a $k_{r}$ device or a single $x$ energy for a $k_{x}$ device, as
an `ideal filter'. For a ballistic device this may be expressed as a
transmission probability function as
\begin{equation}
    \label{ideal_transprob}
    \zeta(E) = \left\{ \begin{array}{cc}
        1 & E = E'\\
        0 & \mbox{elsewhere}
    \end{array} \right.
\end{equation}
where $E_{x}$ would be substituted for $E$ in a $k_{x}$ system.

In a thermoelectric device, inhomogeneous doping or a graded band
structure is required so that Eq. \ref{E0} may be satisfied at every
point in the device for a particular temperature gradient such that
the energy gap between the chemical potential and the transmission
energy is given by \cite{HumphreyPRL2005}
\begin{equation}
    \label{u0}
    E_{0} - \mu_{0}(x) =  \left[ \dfrac{ eV_{OC}} {T_H - T_C} \right] T(x)
\end{equation}
where $V_{OC}$ is the open circuit voltage.

\section{Electronic Efficiency With Ideal Filtering}
\label{ideal_filtering_section}

\begin{figure}
\includegraphics[width=2.5in]{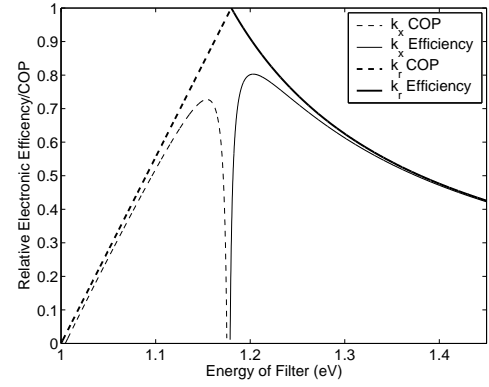}
\caption{\label{Efficiency_Ideal_Filtering} Relative efficiency and
COP of ideally filtered $k_{r}$ and $k_{x}$ devices versus the
energy of the ideal filter. The $k_{x}$ curves are plotted against
the total average energy of electrons leaving the cold reservoir,
$E_{x} + k_{B}T_{C}$. $T_{H} = 300$ K, $T_{C} = 270$ K, $\mu_{H} =
0.98$ eV and $\mu_{C} = 1.00$ eV.}
\end{figure}

Under ideal filtering, as defined in the previous section, Eqs.
\ref{efficiency} and \ref{COP} for the $k_{r}$ device reduce to
\begin{equation}
    \label{kr_ideal_efficiency}
    \eta_{r}^{PG} = \dfrac{\mu_{C} - \mu_{H}}{E_{0} - \mu_{C}}
\end{equation}
for power generation ($E' > E_{0}$) and
\begin{equation}
    \label{kr_ideal_COP}
    \eta_{r}^{R} = \dfrac{E_{0} - \mu_{C}}{\mu_{C} - \mu_{H}}
\end{equation}
for refrigeration ($E' < E_{0}$). The efficiency and COP of ideally
filtered $k_{x}$ and $k_{r}$ systems are plotted in Fig.
\ref{Efficiency_Ideal_Filtering} relative to the Carnot values. When
the filtering energy is $E_{0}$, reversibility and the Carnot
efficiency are achieved for the $k_{r}$ device as shown in Fig.
\ref{Efficiency_Ideal_Filtering}. The energy axis for the $k_{x}$
device shown in Fig. \ref{Efficiency_Ideal_Filtering} is the average
total cold reservoir energy, $E_{x} + k_{B}T_{C}$.

\begin{figure}
\includegraphics[width=2.5in]{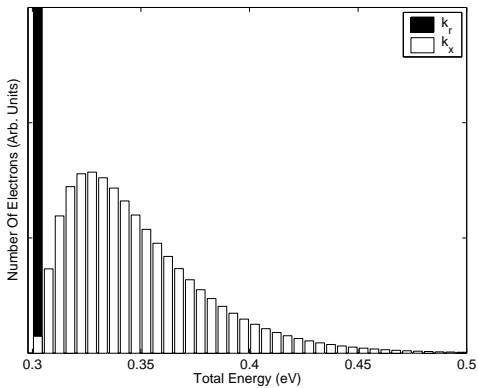}
\caption{\label{kxTotalEnergyDistribution} The total energy
distributions for electrons leaving a reservoir in $k_{x}$ and
$k_{r}$ filtered thermionic devices with a filter of 5 meV at 0.3
eV. The vertical axis has been cut off for clarity of the details in
the $k_{x}$ system values. The number of electrons transmitted in
the $k_{r}$ system is approximately 13 times that of the $k_{x}$
system for these parameters. $T =$ 270 K and $\mu =$ 0.1 eV.}
\end{figure}

For all values of total energy shown in Fig.
\ref{Efficiency_Ideal_Filtering}, the $k_{r}$ device outperforms the
$k_{x}$ device. Importantly, unlike the $k_{r}$ device, the $k_{x}$
filtered thermionic device does not reach the Carnot efficiency for
arbitrary electrochemical potentials and finite barrier heights. The
reason for this is that although momentum in the $x$ direction is
restricted to a single value, the momentum in the $y$ and $z$
directions may take any value, meaning that the energy spectrum has
a finite width and reversibility is not achieved. The distributed
nature of total electron energies for a $k_{x}$ device, even with a
narrow filter, is shown in Fig. \ref{kxTotalEnergyDistribution}. For
$k_{x}$-filtered power generators, this upper bound upon the
electronic efficiency can be obtained analytically in the limit that
($\mu_{C} - \mu_{H})/k_{B}(T_{H} - T_{C}) \gg 1$, in which case
maximum efficiency is obtained when $E'=E_{0}$, where
\begin{equation}
    \label{kx_PG_max_eff}
    \eta_{x}^{PG} = \eta_{C} \left[ 1 + \eta_{C}(k_{B}T_{H} +
    k_{B}T_{C}) / qV \right]^{-1}
\end{equation}
where $\eta_{C}$ is the Carnot efficiency. Given that we have taken
$\mu_{C} > \mu_{H}$, so that $qV$ is positive, it can be seen by
inspection that Eq. \ref{kx_PG_max_eff} always yields an efficiency
less than $\eta_{C}$. For the system analysed in Fig.
\ref{Efficiency_Ideal_Filtering}, Eq. \ref{kx_PG_max_eff} gives a
maximum electronic efficiency for the $k_{x}$ power generator of
80\% of the Carnot limit, in agreement with numerical results. This
constitutes the first main result of the paper, that for finite
barrier heights and electrochemical potential differences, $k_{x}$
filtered thermionic devices are limited to a maximum electronic
efficiency less than the Carnot limit. This means that from the
point of view of maximising electronic efficiency, $k_{r}$ devices
are inherently superior to $k_{x}$ devices.

\section{Electronic Efficiency With Non-Ideal Filtering}
\label{NonIdealFiltering}

The filters considered in Sect. \ref{ideal_filtering_section}
represent an idealized theoretical limit. We now extend our analysis
to non-ideal filters. Firstly, we consider filters of finite width
where all electrons over a certain range of total or $x$ energies
are transmitted. Secondly, we show that a gradual, rather than sharp
switch from zero to full transmission has a significant impact on
the electronic efficiency of thermionic devices. This will
constitute the second main result of the paper.

\subsection{Effect of Finite Filter Width \label{Effect_Of_Finite_Filter_Width}}

\begin{figure}
\includegraphics[width=2.5in]{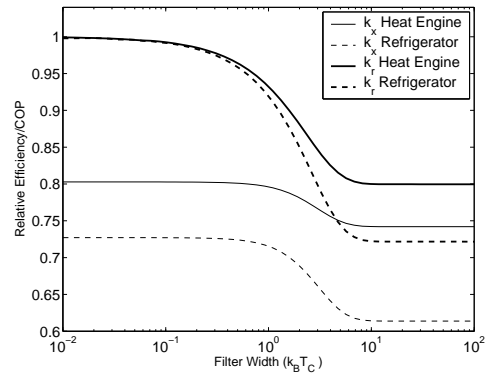}
\caption{\label{Efficiency_Finite_Filter_Width} The maximum relative
efficiency and COP as a fraction of the Carnot limit for $k_{r}$ and
$k_{x}$ filtered thermionic devices versus the width of the filter
from $0.01k_{B}T_{C}$ to $100k_{B}T_{C}$. $T_{H} = 300$ K, $T_{C} =
270$ K, $\mu_{H} = 0.98$ eV and $\mu_{C} = 1.00$ eV.}
\end{figure}

 A filter of finite width
corresponds to a transmission probability of
\begin{equation}
    \label{ideal_transprob}
    \zeta(E) = \left\{ \begin{array}{cc}
        1 & E' < E < E' + \Delta E\\
        0 & \mbox{elsewhere}
    \end{array} \right.
\end{equation}
for a $k_{r}$ system, and where $E_{x}$ would be substituted for $E$
for a $k_{x}$ system. Such a filter might be used, for example, to
approximate a transmission miniband in a superlattice device. For
each filter width examined numerically, the starting energy of the
filter, $E'$, was tuned to find the maximum electronic
efficiency/COP for that width. The results for filters of width
0.01$k_{B}T_{C}$ to 100$k_{B}T_{C}$ are plotted in Fig.
\ref{Efficiency_Finite_Filter_Width}. The filter of 0.1$k_{B}T_{C}$
is narrow enough to effectively perform ideal filtering, reflected
in the fact that the $k_{r}$ electronic efficiency/COP approaches
the the Carnot value for this width and the $k_{x}$ values reach the
maximum values obtained in Fig. \ref{Efficiency_Ideal_Filtering}.
Fig. \ref{Efficiency_Finite_Filter_Width} shows however that we do
not require an ideal filter to achieve an efficiency/COP very close
to the maximum value, as seen in the plateau in all curves. The
$k_{r}$ and $k_{x}$ systems may achieve an efficiency/COP
approximately equal to the maximum value for filter widths of less
than about 0.1$k_{B}T_{C}$ and $k_{B}T_{C}$ respectively. Filter
widths of around these sizes are achievable using practical
semiconductor devices as will be discussed later. As the filter
widths increase beyond these values the efficiency/COP drops and
then plateaus again at a final value. Large filter widths
effectively correspond to the situation where all electrons above
$E'$ are being transmitted. As the distribution function rapidly
converges to zero at high energies, this means that further
increasing filter width has a minimal effect upon the electronic
efficiency.

\begin{figure}
\includegraphics[width=2.5in]{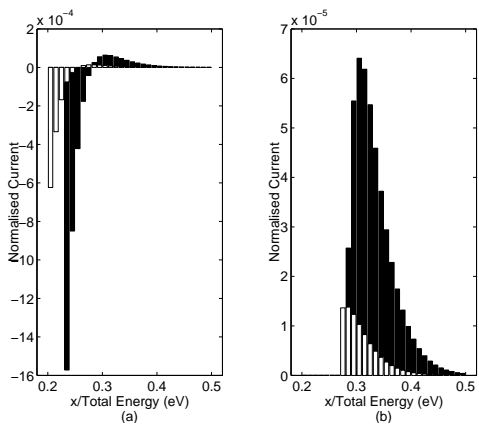}
\caption{\label{Energy_Spectrum} The normalised electron energy
spectrum for net current from the hot to cold reservoir. The filter
width is 0.3~eV and is positioned to achieve maximum efficiency/COP
in each system for (a) refrigeration and (b) power generation in
$k_{r}$ (black) and $k_{x}$ (white) systems. $T_{C}$ = 270~K,
$T_{H}$ = 300~K, and $\mu_{C}$ = 0.1~eV, $\mu_{H}$ = 0.08~eV.}
\end{figure}

Fig. \ref{Energy_Spectrum} shows the energy spectrum of the net
electric current transmitted from the hot to cold reservoir for a
0.3 eV wide filter. Results are normalized by the net number of
electrons with total energy greater than the Fermi energy available
to flow between reservoirs such that the number of electrons in
$i$th energy band are given by
\begin{equation}
    \label{Energy_Spectrum_Equation}
    N_{i} = \dfrac {\int^{E_{i} + \delta i}_{E_{i}} n_{r/x} dE} {| \int^{\infty}_{\mu_{C}} n_{r}
    dE |}.
\end{equation}
It should be noted that the $k_{x}$ energy spectrum in Fig.
\ref{Energy_Spectrum} does not show the total energy spread due to
the unfiltered degrees of freedom, as was shown in Fig.
\ref{kxTotalEnergyDistribution}, since in this case we are
considering the spectrum with regard to filtered $x$ component of
energy only. This illustrates the energy range of the filter for
each system when tuned for maximum electronic efficiency/COP.

Fig. \ref{Energy_Spectrum} shows that there are more electrons being
transmitted for the $k_{r}$ system than with the $k_{x}$ system, an
effect previously pointed out by Vashaee and Shakouri
\cite{VashaeeJAP2004, VashaeePRL2004}, which results in greater
power in a $k_{r}$ device. The calculations presented here show that
the difference in the energy spread of electrons in $k_{x}$ and
$k_{r}$ filtered devices also gives an increase in the electronic
efficiency for $k_{r}$ devices due to a greater concentration of
electrons with energies around $E_{0}$. For both refrigeration and
power generation, the filters will be positioned such that electrons
with energy $E_{0}$ are included. Since when $E> E_{0}$ the net
energy-resolved current produces power, the lower edge of the
$k_{r}$ power generator filter will always be at $E_{0}$. For the
$k_{x}$ system this lower edge is shifted to lower $x$ energy due to
the additional energy contribution in the two other spatial degrees
of freedom. Energy-resolved current in the energy range $\mu_C < E <
E_{0}$ refrigerates the cold reservoir and the lower edge of the
filter is therefore shifted to this region in Fig.
\ref{Energy_Spectrum}(a). Since there are more electrons at higher
energies, however, current flow in the region $E > E_{0}$ generates
power and a trade off occurs when positioning the filter for maximum
COP. Again, the $x$ energy of the $k_{x}$ filter is lower than the
total energy of the $k_{r}$ filter due to the unfiltered energy
contributions.

\subsection{Transmission Probabilities With Finite Slopes}

Thus far we have considered only the case where there is a sharp
transition from zero to full transmission of electrons. In this
section we consider the effect upon the electronic efficiency of a
gradual transition, which more closely resembles the shape of the
transmission probability in practical devices. We begin by using two
convenient `artificial' transmission probabilities, the slope of
which can be easily varied. The first, a Gaussian peak which might
approximate the transmission probability of a resonance, is given by
\begin{equation}
    \label{Gaussian}
    \zeta(E) = \exp \left( - \dfrac {(E - E_{c})^{2}} {w} \right)
\end{equation}
where $E_{c}$ defines the center energy of the peak and $w$ is a
width parameter which is used to vary the sharpness of the slope.
$E_{x}$ would be substituted for $E$ for a $k_{x}$ device.

\begin{figure}
\includegraphics[width=2.5in]{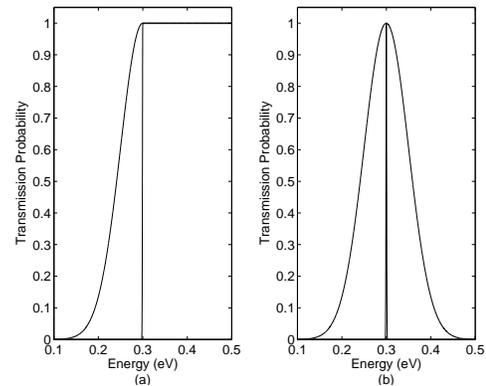}
\caption{\label{Effect_Of_Finite_Slope_Trans_Probs} Artificial
transmission probabilities generated using a (a) Gaussian function
and (b) a half-Gaussian function. The two extremes of width
parameter are shown, $w$~=~0.1 (slowly rising) and $w$~=~0.00001
(sharply-rising).}
\end{figure}

\begin{figure}
\includegraphics[width=3.2in]{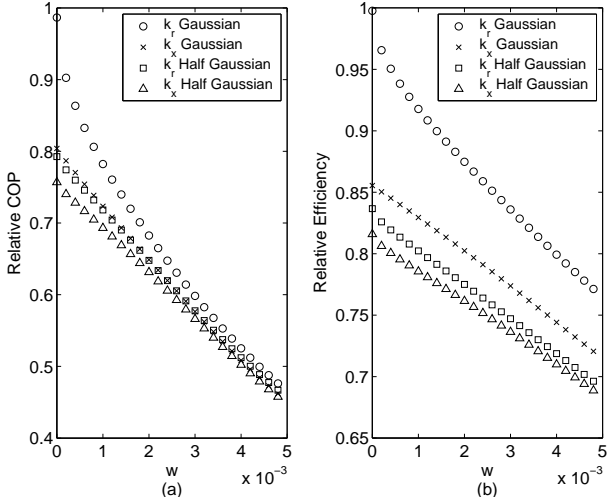}
\caption{\label{Effect_Of_Finite_Slope} The efficiency/COP relative
to the Carnot value of a $k_{r}$ Gaussian, $k_{x}$ Gaussian, $k_{r}$
half-Gaussian and $k_{x}$ half-Gaussian for (a) refrigerator systems
($T_{H}$ = 300~K and $T_{C}$ = 270~K) and (b) heat engine systems
($T_{H}$ = 900~K and $T_{C}$ = 300~K) versus width parameter with
$\mu_{C}$ = 0.1~eV.}
\end{figure}

The second artificial transmission probability considered is a
`half-Gaussian' intended to approximate the transmission probability
of a single barrier of finite width. This is given by Eq.
\ref{Gaussian} for $E \leq E_{C}$ and is equal to one for $E >
E_{C}$. The sharpness of the Gaussian and half-Gaussian transmission
probabilities were varied between $w = 10^{-5}$, corresponding to an
ideal filter or perfectly sharp single barrier transmission
probability, and $w = 0.1$. The transmission probabilities
associated with these extreme values are shown in
\ref{Effect_Of_Finite_Slope_Trans_Probs}(a) and (b). The system bias
was tuned for each transmission probability for maximum electronic
efficiency/COP. Fig. \ref{Effect_Of_Finite_Slope}(a) shows the COPs
associated with a room temperature refrigerator and Fig.
\ref{Effect_Of_Finite_Slope}(b) shows the electronic efficiencies of
a heat engine operating at higher temperature.

Since all electrons of energy other than $E_{0}$ reduce the
electronic efficiency we expect the sharpest peak in Figs.
\ref{Effect_Of_Finite_Slope}(a) and (b) to yield the highest
efficiency/COP, and this is confirmed by the numerical results. The
most interesting result, however, is that the electronic efficiency
of the half-Gaussian transmission probability is very strongly
dependent upon how sharply the transmission rises from zero to
unity. A smooth rise in the transmission probability lowers the
electronic efficiency for the same physical reason that a $k_{x}$
filtered device has a lower electronic efficiency than a $k_{r}$
filtered device. Net current flow is from the hot to cold reservoir
for electron energies just above $E_{0}$, generating power with
efficiency approaching the Carnot limit. Conversely, net current
flow is from the cold to the hot reservior for electron energies
below $E_{0}$ and above $\mu_{C}$, absorbing power and refrigerating
the cold reservoir, while the net current transmitted at energies
below $\mu_{C}$ both absorb power and heat the cold reservoir. This
means that whenever the energy spectrum of transmitted electrons
rises slowly to its peak value there is an efficiency lowering trade
off which occurs between transmitting the maximum number of
electrons with energies near $E_{0}$, which refrigerate or generate
power with Carnot efficiency, and minimising the number of electrons
transmitted in the range $E < E_{0}$ for power generation or in the
range $E < \mu_{C}$ and $E > E_{0}$ for refrigeration. As the
transmission probabilities become less sharp, the performance
difference between the $k_{x}$ and $k_{r}$ and Gaussian and
half-Gaussian systems becomes less significant.

So far we have established the two criteria for maximum electronic
efficiency in thermionic power generators and refrigerators.
Firstly, we have shown that the narrower the energy spectrum the
higher the electronic efficiency. However, in general a gain in
electronic efficiency via this mechanism is obtained at the expense
of the power of the device. The second criterion is that the sharper
the transition from zero to peak value in the energy spectrum, the
higher the electronic efficiency. This second method offers the
significant advantage of improving the electronic efficiency without
sacrificing power through the use of a narrow filter. The maximum
power achievable is also greater with a sharply-rising transmission
probability if the barrier height is optimised
\cite{HumphreyJPD2005}. In the next section we analyse design
considerations for thermionic devices considering both electronic
efficiency and power.

\section{Design Considerations For Achieving High Electronic
Efficiency In Practical Devices}

\subsection{Ballistic Devices}

Semiconductor-based devices, including superlattices, may be
specifically designed to achieve the desired energy spectrum
features in $k_{x}$ devices. Filter widths around those required for
achieve near-maximum electronic efficiency, as discussed in
Sect.\ref{Effect_Of_Finite_Filter_Width}, may be achieved using a
variably-spaced superlattice energy filter (VSSEF) as proposed by
Summers, Brennan and Gaylord \cite{SummersAPL1986,GaylordAPL1988}.
Such devices consist of alternating semiconductor layers with
barrier and well widths chosen such that energy levels in the wells
are closely aligned under bias. Tunneling through a simpler
multibarrier structure may also suffice. Similarly, a miniband in
the transmission probability for a superlattice might be used as a
narrower filter compared with complete transmission above the
barrier energy. Quantum dot structures \cite{BryllertAPL2003} or
normal-insulating-superconductor junction (NIS) devices
\cite{EdwardsPRB1995} can also achieve narrow electron transmission
bands and may be used for refrigeration at cryogenic temperatures.
Relatively narrow energy electron emission peaks from carbon
nanotubes have been reported which may be of use in a vacuum based
device \cite{FransenASP1999}. Since the DOS in the reservoirs fixes
the number of electrons available for transport in a certain energy
range in ballistic devices, the reduction of power in narrow
transmission probability devices, with only modest gains in
electronic efficiency, is expected to be undesirable in the presence
phonon heat leaks.

It is likely that the best way to simultaneously achieve high
electronic efficiency and high power in a ballistic device is to
design the structure such that the transmission probability rises
sharply from zero to one and remains close to unity beyond this.
Whilst the most obvious way to achieve a transmission probability of
this nature is to utilise a single barrier with a width as large
possible (but less than the mean free path of electrons) here we
show that an array of thin barriers can also be used to engineer a
transmission probability that rises sharply from zero to unity.

\begin{figure}
\includegraphics[width=2.5in]{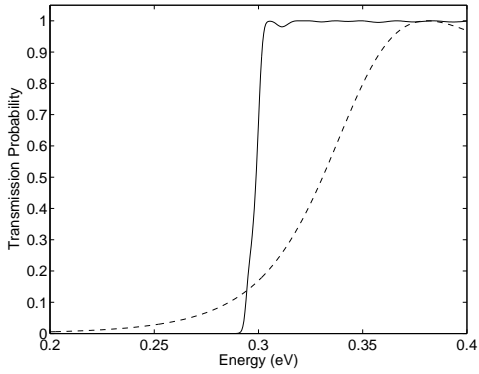}
\caption{\label{Sharpness_Single_Barrier} The transmission
probability of a 10-nm (dashed line) and 100-nm (solid line) rounded
single 0.3-eV barrier systems under no applied bias. Effective mass
is $0.067m_{e}$.}
\end{figure}

\begin{figure}
\includegraphics[width=2.5in]{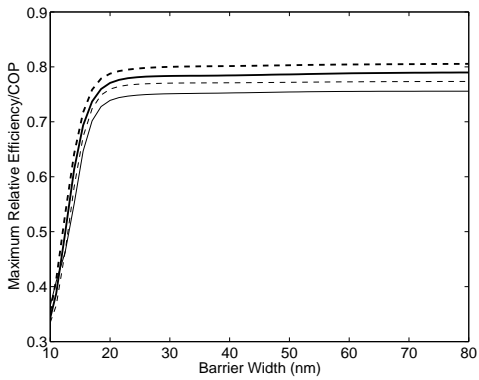}
\caption{\label{Efficiency_Versus_Barrier_Width} The relative
electronic efficiency and COP of single barrier $k_{r}$ heat engine
(thick-dashed line), $k_{x}$ refrigerator (thick-solid line),
$k_{x}$ heat engine (thin-dashed line) and $k_{x}$ refrigerator
(thin-solid line) devices versus the width of the barrier. $T_{C} =
270$~K, $T_{H} = 300$~K, $\mu_{C} = 0.1$~eV, barrier height of
0.3~eV and effective mass of 0.067$m_{e}$.}
\end{figure}

The transmission probabilities and associated efficiencies/COPs for
single rounded barriers of various widths have been calculated. The
transmission probabilities were calculated by obtaining a numerical
solution to the time-independent Schr\"odinger equation based on
Airy functions \cite{Davies1998,BrennanJAP1987}. Fig.
\ref{Sharpness_Single_Barrier} shows the significant difference in
sharpness between a 10-nm and 100-nm barrier. We see in Fig.
\ref{Efficiency_Versus_Barrier_Width}, as expected, the wider
barriers with sharper transmission probabilities give the highest
efficiencies/COPs approaching the maximum value, in this case, at a
width of around 35 nm. Beyond this width, there is little to be
gained in terms of electronic efficiency, although phonon mediated
heat leaks continue to be reduced with increasing barrier width.

From another point of view, wide barriers might be undesirable.
Devices with greater interface density may reduce thermal
conductivity as a result of interface scattering and phonon miniband
formation \cite{CahillAPR2003}. Here we consider a device where
multiple barriers are traversed in an electron mean free path. This
allows quantum mechanical effects to be utilised to achieve high
electronic efficiency using narrow barriers which give low
electronic efficiency when used individually.

\begin{figure}
\includegraphics[width=2.5in]{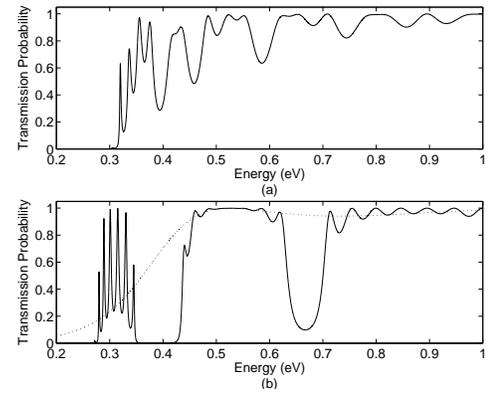}
\caption{\label{Multibarrier_Transprobs} Transmission probabilities
calculated for (a) a two 20-nm barrier system with 20-nm well and
(b) an eight 5-nm barrier system with 5-nm wells (solid) and the
transmission probability for a single 5-nm barrier (dotted) for
comparison. $T_{C} = 270$~K, $T_{H} = 300$~K, $\mu_{C} = 0.1$~eV,
barrier height of 0.3~eV and effective mass of 0.067$m_{e}$.}
\end{figure}

Multiple narrow barriers over a distance of the order of the
electron mean free path may be used to give a transmission
probability that is as sharp as if the electrons were traversing a
single wide barrier. Figs. \ref{Multibarrier_Transprobs}(a) and (b)
show the transmission probabilities calculated for two-barrier and
eight-barrier systems respectively as well as the very smoothly
rising transmission probability for a single 5-nm barrier for
comparison. The efficiencies/COPs achieved are within 3\% of those
of a wide single barrier. Thus, high electronic efficiencies may be
achieved, whilst allowing the flexible use of narrower barriers.

\begin{figure}
\includegraphics[width=2.5in]{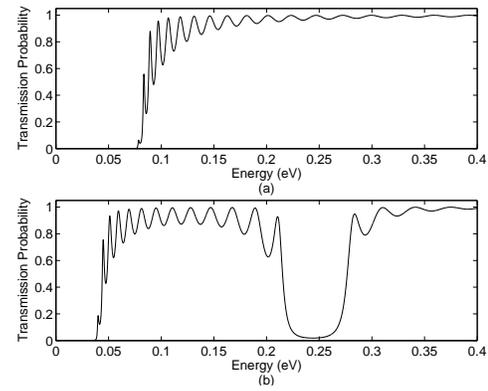}
\caption{\label{Many_Barrier_Transprobs} Transmission probabilities
calculated for: (a) An 80-nm single barrier with maximum $k_{x}$
refrigerator COP of 0.59 when $\mu_{C} = 0$~eV and 0.48 when
$\mu_{C} = 0$~eV, (b) A 15 3-nm barrier system with 3-nm wells with
$k_{x}$ refrigerator COPs of 0.48 when $\mu_{C} = 0$~eV. $T_{C} =
270$~K, $T_{H} = 300$~K, barrier height of 0.075~eV and effective
mass of 0.05$m_{e}$.}
\end{figure}

The `turn-on' transmission energy for a device with many thin
barriers may be shifted to lower energy as shown in Fig.
\ref{Many_Barrier_Transprobs}(a) and (b). If the chemical potential
remains constant, this lowering of the turn-on energy may result in
a decrease in efficiency and increase in power compared to a wider
single barrier due to the decrease in the work function, $\phi =
E_{turnon}-\mu$. With the same work functions, which may be achieved
by altering the chemical potentials as detailed in Figs.
\ref{Many_Barrier_Transprobs}(a) and (b), the wide single barrier
and many thin barrier systems achieve approximately the same
electronic efficiency.

We do not expect a dramatic change in dependence of device behaviour
on the electron energy spectrum as the total length of the device
increases beyond an electron mean free path. Since the probability
of an electron traveling distance $L$ without suffering a collision
is given by~\cite{Ashcroft1976}
\begin{equation}
    \label{Probability_Of_Collision}
    P = \exp(-\lambda / L)
\end{equation}
the ballistic and diffusive formalisms, Eqs. \ref{ballistic} and
\ref{diffusive}, may be combined to show that the electrical current
will be given by~\cite{HumphreyJAP2005}
\begin{equation}
    \label{Ballistic_Diffusive_Current}
    J = \iiint q \Delta f \left( D_{r} v^r_{x} \zeta P +  \dfrac{\lambda}{L} D_{l} v^l_{x} [1 - P] \right)
    d\textbf{k}
\end{equation}
if $L \approx \lambda$. As the number of barriers increases and the
overall length of the device becomes significantly greater than the
electron mean free path, the ballistic term becomes small so that
transport is accurately described using the diffusive formalism, as
discussed in the next section.

\subsection{Diffusive Devices}

\begin{figure}
\includegraphics[width=2.5in]{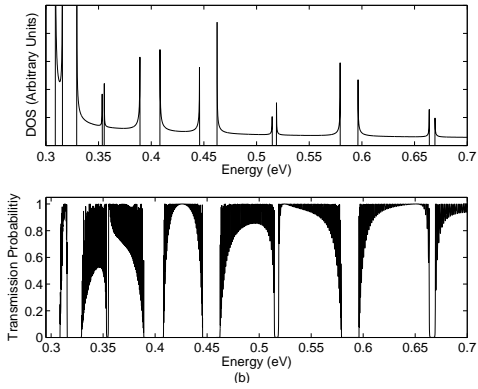}
\caption{\label{DOS} (a) The local DOS for an infinite series of
20-nm barriers separated by 20 nm and (b) the transmission
probability for fifty 20-nm barriers separated by 20 nm with a
barrier height of 0.3~eV and effective mass of 0.067$m_{e}$.}
\end{figure}

As was discussed earlier, the energy spectrum in a diffusive device
is determined by the product of the local DOS, the velocity squared
and the relaxation time at constant temperatures/chemical
potentials. The local DOS of an infinite superlattice may be
determined using the Kronig-Penney model
\cite{Davies1998,LinPRB2003}. Fig. \ref{DOS} shows the DOS
calculated for a many-barrier system and the calculated transmission
probability showing the clear relationship between the two. In a
pure ballistic device, since the DOS and velocity are fixed by the
reservoirs, a sharp electron energy spectrum is achieved via a sharp
transmission probability. In diffusive devices, both the DOS and
electron velocity may change as the device structure changes and it
may be a difficult optimisation problem to design a structure where
their product changes sharply from its minimum to maximum value as a
function of energy \cite{ShakouriPersonalCommunication}.

Mahan and Sofo have shown that the ideal transport distribution
function for a thermoelectric device may be achieved with a delta
function DOS \cite{MahanNAS1996}. Humphrey and Linke showed that the
Carnot efficiency may be achieved in thermoelectric devices
utilising a delta function DOS and a graded device structure or
inhomogeneous doping \cite{HumphreyPRL2005}. Their results are
analogous to the results presented earlier in this paper where it
was shown that the ideal transmission probability for a ballistic
device was one which allowed the transmission of only a very narrow
energy range of electrons. The results presented in this paper
suggest that not only is the width of the energy spectrum important,
but also whether it rises rapidly from zero to its maximum value. In
practical devices with loss mechanisms such as phonon heat leaks,
the magnitude of the energy spectrum also becomes important to the
efficiency as the conductivity is given by the integral of the
energy spectrum and occupation of states. Hicks and Dresselhaus have
pointed out that the magnitude of the DOS can be increased by using
structures of lower dimensionality, potentially increasing the power
factor \cite{HicksPRB1993a,HicksPRB1993b}. We also note that the DOS
is sharper for lower dimensional systems compared to bulk materials,
which may result in an improved energy spectrum.

\section{Experimentally Observable Properties Related To
Electronic Efficiency}

The presence of phonon heat leaks and contact resistance in
solid-state thermionic devices makes direct measurements of the
electronic efficiency difficult. Here we discuss experimental
properties which may be measured to provide an indication of the
shape of the electron energy spectrum and though this, electronic
efficiency.
\begin{figure}
\includegraphics[width=2.5in]{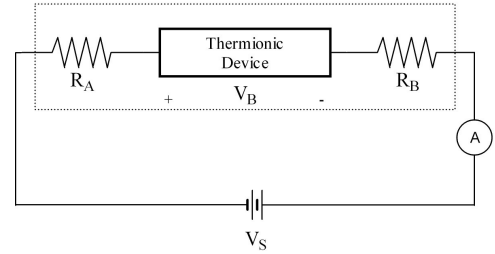}
\caption{\label{TI_Circuit} A thermionic circuit showing contact
resistances $R_{A}$ and $R_{B}$, source voltage $V_{S}$ and barrier
voltage $V_{B}$.}
\end{figure}
\begin{figure}
\includegraphics[width=2.5in]{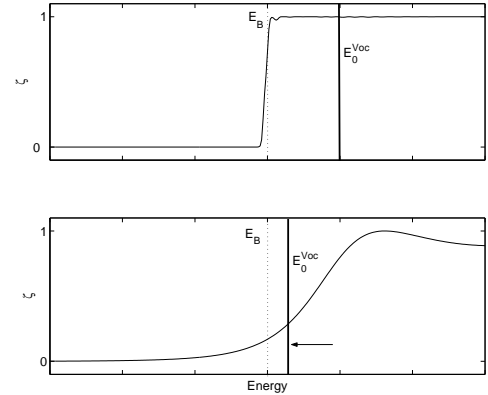}
\caption{\label{Sharpness_Indicator_Fig} The transmission
probabilities for (a) a wide  and (b) a narrow single barrier.
$E_{0}^{V_{0C}}$ is indicated by the dashed line and the barrier
energy, $E_{B}$ by the dotted line.}
\end{figure}

\begin{figure}
\includegraphics[width=2.5in]{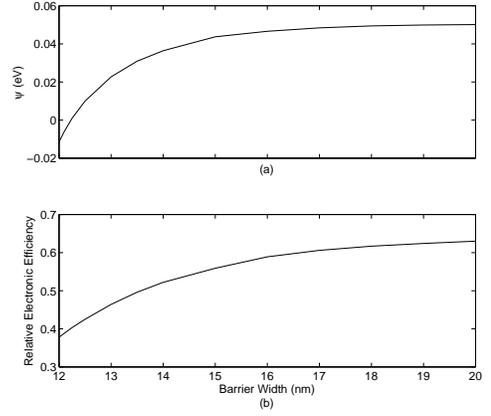}
\caption{\label{Experimental_Measures} The (a) maximum relative
electronic efficiency and (b) sharpness indicator versus single
barrier width. The cold reservoir chemical potential has been tuned
so that the Seebeck coefficient is constant for each barrier width.
$T_{C} = 270$~K, $T_{H} = 300$~K, barrier height of 0.3~eV and
effective mass of 0.067$m_{e}$.}
\end{figure}

The I-V characteristics of a thermionic device are dependent on the
voltage across the barrier system, $V_{B}$, as shown on Fig.
\ref{TI_Circuit}. $V_{B}$ may be determined from the supplied
voltage, $V_{S}$, and measured current, $I$, as
\begin{equation}
    \label{Vb}
    V^{B} = V^{S} + I(R_{A} + R_{B})
\end{equation}
where, $R_{A}$ and $R_{B}$ are contact resistances, when the device
is generating power. As the bias is increased, the net electrical
current decreases and reaches zero at the open circuit voltage
$V^{B}_{OC}$, from which the effective Seebeck coefficient may be
calculated as
\begin{equation}
    \label{Seebeck}
    S = \dfrac{V^B_{OC}}{T_{H}-T_{C}}.
\end{equation}
The energy-specific equilibrium energy may be calculated at open
circuit voltage as
\begin{equation}
    \label{EoVoc}
    E_{0}^{V_{OC}} = \mu_{C} + V_{OC}^{B}\dfrac{T_{C}}{T_{H}-T_{H}}
    = \mu_{C} + S T_{C}
\end{equation}
and is linearly related to the Seebeck coefficient. $E_{0}^{V_{OC}}$
is the energy where energy-resolved currents above and below it are
equal, giving zero net current. For a sharply-rising transmission
probability, $E_{0}^{V_{OC}}$ would be be positioned as shown in
Fig. \ref{Sharpness_Indicator_Fig}(a) above the barrier energy. If
we have another system where electrons with energies lower than the
barrier energy are being transmitted without significant change to
the high-energy details, for example through decreasing the barrier
width, $E_{0}^{V_{OC}}$ is shifted to lower energy as shown in Fig.
\ref{Sharpness_Indicator_Fig}(b). Measuring this relative to a
convenient energy, say the barrier energy, $E_{B}$, provides a
convenient sharpness indication for the transmission probability,
\begin{equation}
    \psi = S T_{C} - \phi.
\end{equation}
The `turn-on' energy for a multibarrier system may be shifted to
lower energy, in which case, the sharpness indicator should be
measured relative to this `effective' barrier height, which might be
calculated using the Kronig-Penney model, as discussed previously. A
higher sharpness indicator is desirable, indicating a sharper
transmission probability and therefore higher expected electronic
efficiency/COP. The sharpness indicator has the advantage over the
Seebeck coefficient of being less dependent on the chemical
potential/barrier height and more so on the sharpness of the energy
spectrum, as shown in Fig. \ref{Experimental_Measures}. Here, the
chemical potentials for a number of single barrier transmission
probabilities have been varied to give a constant Seebeck
coefficient as the barrier width and transmission probability
sharpness change. Fig. \ref{Experimental_Measures} shows the
electronic efficiency varies significantly in this example. Whilst
the Seebeck coefficient remains constant, the sharpness indicator,
$\psi$, increases as the electronic efficiency increases.

\section{Conclusions}

It has been shown the the nature of the electron-energy spectrum has
a significant impact on the performance of thermionic and
thermoelectric devices. The limiting efficiency of finite barrier
height devices was achieved when electrons of a single energy only
are transmitted. Whilst $k_{r}$ devices achieve reversibility when
the transmission energy was equal to $E_{0}$, $k_{x}$ devices do not
due to the finite energy spread associated with the two unfiltered
degrees of freedom. For systems with finite-width rectangular
transmission probabilities, electronic efficiency was close to the
maximum value for filter widths less than $k_{B}T$, but decreases as
the range of transmitted electron energies increases, reaching a
steady value as the filter width increases beyond a few $k_{B}T$.
Our most important result was that an increase in the sharpness of
the rise in electron energy spectrum significantly increases
electronic efficiency. Improving the electronic efficiency by
increasing the sharpness of the transmission probability may also
increase the maximum power. We have shown that sharp transmission
probabilities may be achieved using wide single barriers or
carefully arranged multiple barriers.

Since, in the diffusive formalism, used to describe thermoelectric
devices, the product of the local electron group velocity and the
local DOS plays the same role as the product of the reservoir DOS,
reservoir velocity and the transmission probability in the ballistic
formalism for mean free path length devices, the results presented
here showing the benefit of sharply-rising energy spectra on
electronic efficiency and power are expected to be relevant to
thermoelectric devices.

Finally, the sharpness indicator, $\psi$, was suggested as an
experimental measure providing an indication of the sharpness of the
rise in the energy spectrum of a ballistic device and its electronic
efficiency and was shown to be superior for this purpose to the
Seebeck coefficient alone.

\section{Acknowledgements}
MO'D is supported by the Australian Research Council. TH is
supported by the Australian Research Council and funding from ONR
MURI. The authors acknowledge helpful discussions with Ali Shakouri
and Heiner Linke.

\newpage 


\end{document}